\documentclass[utf8,english]{jetp}

\twocolumn

\begin{document}
\English

\title{Axion--neutrino interactions in seesaw models and astrophysical probes}

\author{P.}{Kivokurtseva}
\email{kivokurtceva.pi19@physics.msu.ru}
\affiliation{Institute for Nuclear Research of the Russian Academy of Sciences,\\
60th October Anniversary Prospect 7a, Moscow 117312, Russia}

\abstract{
We study axion--neutrino interactions in neutrino-mass extensions of the Standard Model, focusing on the Type-I and inverse seesaw.
In benchmark scenarios where the effective ALP charge of the light neutrino is comparable to the electron coefficient, the coupling $g_{a\nu}$ can be related to the axion scale and constrained using existing limits on axion couplings to electrons.
We then estimate the impact of axion-mediated scattering on neutrino propagation in two benchmarks: resonant interactions with the C$\nu$B and scattering on axion dark matter.
In the parameter space allowed by current bounds, the resulting optical depths are extremely small, implying no observable signatures with present sensitivities.
}

\maketitle

%\tableofcontents

\section{Introduction}
\label{sec:intro}

Axions and axion-like particles (ALPs) are hypothetical pseudo-Nambu–Goldstone bosons that arise in a variety of extensions of the Standard Model, most famously in the Peccei–Quinn solution to the strong-CP problem ~\cite{PhysRevLett.38.1440,PhysRevD.16.1791}. Beyond their well-known two-photon coupling that enables photon–axion mixing in electromagnetic fields, axions/ALPs can also interact with fermions. In many theoretically well-motivated constructions these couplings scale with the fermion mass, schematically
\begin{equation}
\mathcal{L} \supset i\, g_{a f}\, a\, \bar{f}\gamma_5 f, 
\qquad
g_{a f} \propto \frac{m_f}{f_a},\end{equation}
where $a$ is the axion field, $f$ a Standard Model fermion, and $f_a$ the Peccei–Quinn symmetry-breaking scale (or a generic UV (ultraviolet) scale in ALP scenarios). While the axion–photon interaction has been extensively exploited in laboratory experiments and astrophysical \\ searches \cite{Raffelt_2024}, the axion–lepton sector is currently constrained less tightly and could provide complementary discovery channels.

Of particular interest is the axion–neutrino interaction. Neutrino oscillations demonstrate that neutrinos are massive, implying that an axion–neutrino coupling $g_{a\nu}$ is generically allowed in mass-proportional frameworks. 

Several recent works have proposed scenarios that could shed light on axion-neutrino interactions. One example is \cite{Deppisch_2025}, which probes axion-neutrino couplings in a cosmological context. However, the analysis adopts rather unusual values (several TeV) of the axion decay constant $f_a$. Such choices of $f_a$ typically imply large $g_{a\gamma\gamma}$ that are already tightly constrained by photon-axion mixing. The cosmological viability of the parameter region discussed in \cite{Deppisch_2025} therefore either requires non-standard model-building (e.g., photophobic ALPs or alignment mechanisms that suppress $g_{a\gamma\gamma}$ independently of $f_a$) or a careful reassessment of existing photon-coupling limits.

On the astrophysical side, Ref.~\cite{Murase:2019xqi} proposed a time-domain search strategy for new neutrino interactions based on propagation-induced delays (“neutrino echoes”) from transient multimessenger sources. This idea provides a clean handle complementary to spectral probes and can be adapted to axion-mediated channels. A related axion-neutrino scenario was explored in Ref.~\cite{Reynoso_2022}, where high-energy neutrino propagation through ultralight ALP dark matter was studied and the importance of the assumed mass ordering/flavor structure was emphasized.

In this work we take a complementary approach: rather than postulating an effective $g_{a\nu}$ in isolation, we parameterize the ALP coupling to the neutrino mass eigenstates in seesaw-inspired models and identify the benchmark assumptions under which bounds on $g_{ae}$ can be translated into bounds on $g_{a\nu}$. We then assess, at the order-of-magnitude level, the discovery prospects in representative astrophysical benchmarks. 

The paper is organized as follows. In Sec.~\ref{sec:mod} we discuss minimal SM extensions that give rise to axion-neutrino interactions. In Sec.~\ref{sec:data} we review representative constraints on $g_{ae}$ and translate them into bounds on the neutrino-axion coupling. In Sec.~\ref{sec:flux} we estimate the optical depths in two illustrative astrophysical scenarios. We briefly discuss our results in Sec.~\ref{sec:dis}.

\section{Model overview}
\label{sec:mod}
 In this section we briefly outline how neutrino–axion interactions can emerge in two minimal frameworks: the Type-I seesaw \cite{PhysRevLett.44.912} and the inverse seesaw \cite{Mohapatra:1986bd}.

Let us consider a theory of neutrino–ALP interactions. As a first step, we consider a minimal extension of the Standard Model that generates neutrino masses through the seesaw mechanism. To account for at least two nondegenerate
active-neutrino masses—established by multiple neutrino-oscillation experiments—it is
sufficient to introduce at least two right-handed neutrinos \(N_{iR}\) (\(i=1,2\)).
 For simplicity, in our analytic estimates we work with a
single right-handed neutrino, denoted $N_R$. One-generation setup serves as an
order-of-magnitude proxy, and the parametric scaling of our results straightforwardly
generalizes to the multi-generation case. We also introduce a pseudoscalar axion-like particle (ALP) $a$ , which couples to the neutrino sector through the right-handed neutrinos $N_R$. In addition, we allow for generic ALP couplings to other SM fields,
encoded in $\mathcal{L}_{\rm all}$; in particular, a nonzero electron coupling $g_{ae}$ will be used below
to translate existing laboratory and stellar limits into constraints on the overall ALP scale and,
consequently, on the effective axion-neutrino coupling.
The low-energy Lagrangian relevant for neutrino masses and ALP
interactions can be written as
\begin{equation}
\begin{split}
\mathcal{L} &= \mathcal{L}_{\text{SM}}
+ i\bar N_{R}\partial N_{R}
- (Y_\nu)_{\alpha}\bar L_{\alpha}\tilde H N_{R} \\
&\quad - \frac{1}{2}(\mathcal{M}_R)\overline{N_{R}^{c}} N_{R}
+ \mathcal{L}_{aNN} + \mathcal{L}_{ae}
+ \text{h.c.}.
\end{split}
\end{equation}
where \(L_{\alpha}=(\nu_{\alpha L},\,\ell_{\alpha L})^{\top}\) is the SM lepton doublet,
\(H=(H^{+},H^{0})^{\top}\) is the Higgs doublet, \(\alpha=e,\mu,\tau\), and
\(\tilde H \equiv i \sigma_{2} H^{\ast}\). Here the seesaw mass terms are renormalizable, while $\mathcal L_{aNN}$ and $\mathcal L_{ae}$ are understood as effective ALP interactions suppressed by the scale $f_a$.
The term \(\mathcal{L}_{aNN}\) encodes the ALP–neutrino interaction and may be written as: 
\begin{align}
    \mathcal{L}_{aNN} 
    = \frac{C_{\nu}}{f_a}(\partial_\mu a) 
      \Bar{N}_{R} \gamma^\mu \gamma_5 N_{R}
    = - \frac{2i C_{\nu}}{f_a} m_{N_{R}} a 
      \Bar{N}_{R} \gamma_5 N_{R}, 
\label{eq:LNN}
\end{align}

The electron coupling used below should be understood as a low-energy
effective interaction. With the minimal seesaw field content specified
above, a direct pseudoscalar ALP coupling to charged leptons is not a new
renormalizable operator. A gauge-invariant EFT parametrization before
electroweak symmetry breaking is
\begin{equation}
    \mathcal L_{a\ell}
    =
    \frac{\partial_\mu a}{f_a}
    \left[
    C_L\,\bar L\gamma^\mu L
    +
    C_R\,\bar e_R\gamma^\mu e_R
    \right],
    \label{eq:lepton_eft}
\end{equation}
where $L=(\nu_L,e_L)^T$. After electroweak symmetry breaking, the axial
electron current gives
\begin{equation}
    \mathcal L_{aee}
    =
    -i\frac{(C_R-C_L)m_e}{f_a}
    a\bar e\gamma_5 e
    \equiv
    -i g_{ae}a\bar e\gamma_5 e .
    \label{eq:aee_eft}
\end{equation}
In the notation used below,
\begin{equation}
    2C_e\equiv C_R-C_L,
    \qquad
    g_{ae}=\frac{2C_em_e}{f_a}.
    \label{eq:gae_def}
\end{equation}
This EFT form also makes clear that a coupling to the left-handed lepton
doublet induces correlated interactions with both charged leptons and
neutrinos. In this work we do not perform a full gauge-invariant ALP EFT
analysis; instead, we use the constrained low-energy coupling $g_{ae}$ to
set the effective scale and keep the light-neutrino coupling in terms of
the model-dependent parameter $\kappa_i$ defined below. Loop-induced
couplings to electroweak gauge bosons, including photons, may provide
additional constraints in generic ALP models~\cite{Bonilla_2024}; our
bounds should therefore be interpreted as benchmark estimates based on
$g_{ae}$ alone.

After the electroweak symmetry breaking (EWSB), the neutral-fermion mass matrix in the
\((\nu_L,\; N_R^{\,c})\) basis takes the form
\begin{align}
    \mathcal{M}_\nu =
    \begin{pmatrix}
      0               & \mathcal{M}_D \\
      \mathcal{M}_D^{\top} & \mathcal{M}_R
    \end{pmatrix},
\end{align}
where \(\mathcal{M}_D = Y_\nu v/\sqrt{2}\) is the Dirac mass matrix and
\(\mathcal{M}_R\) is the Majorana mass matrix of the right-handed neutrino(s).
Diagonalizing with a unitary matrix \(U\),
\begin{align}
    U^\dagger \mathcal{M}_\nu U^{\ast}
    = \begin{pmatrix}
        m_\nu & 0 \\
        0     & M_R
      \end{pmatrix},
\end{align}
where \(m_\nu\) and \(M_R\)  are diagonal mass matrices.

In the seesaw limit $|\mathcal{M}_D|\ll |\mathcal{M}_R|$ the active-sterile mixing is controlled by
\begin{align}
\Theta \equiv \mathcal{M}_D\,\mathcal{M}_R^{-1}\,,
\end{align}
so that the full diagonalization matrix can be written in a $2\times2$ block form,
\begin{align}
U =
\begin{pmatrix}
U_{\nu\nu} & U_{\nu N}\\
U_{N\nu} & U_{NN}
\end{pmatrix},
\qquad
U_{N\nu} \simeq \Theta,
\qquad
U_{\nu\nu}\simeq \mathbf{1} + \mathcal{O}(\Theta^2).
\end{align}
We will consistently retain only the leading terms in $\Theta$, implying
$|U_{\nu\nu}|^2=\mathcal{O}(1)$ while $|U_{\nu N}|^2\simeq |U_{N\nu}|^2=\mathcal{O}(\Theta^2)$.
For the partial widths of the relevant
two-body decays we get:

\begin{align}
    \Gamma(a \to \nu \nu)
 = \frac{m_N^2m_a |U_{\nu N}|^4}{2\pi f_a^2}\sqrt{1-\frac{4m_{\nu}^2}{m_a^2}}
\end{align}

Integrating out the heavy state leads to an effective interaction for the light neutrinos.
At leading order in $\Theta$, this EFT (effective field theory) reproduces the same $a\to\nu\nu$ amplitude as the full theory, up to higher-order corrections in $\Theta$.

A similar result is obtained if we instead employ the inverse seesaw mechanism \cite{Mohapatra:1986bd} for mass generation. Compared to the Type-I seesaw, the inverse seesaw (ISS) introduces an additional set
of SM-singlet neutral fermion $S_{R}$. The neutral-lepton field
content is therefore \\ $(\nu_L, N_{R}, S_{R})$.
After electroweak symmetry breaking, the mass Lagrangian reads
\begin{equation}
\mathcal{L}_{\rm mass}
= -\overline{N_{R}} m_D \nu_L
  -\overline{S_{R}} M N_{R}^{c}
  - \tfrac{1}{2} \overline{S_{R}} \mu S_{R}^{c}
  + \text{h.c.},
\end{equation}
where $m_D$ is a Dirac mass matrix generated by Yukawa couplings to the Higgs field,
$M$ is a Dirac mass term linking $N_R$ and $S_R$, and $\mu$ is a small Majorana
mass term for $S_R$ that softly breaks lepton number.

In the basis $(\nu_L, N_{R}^{c}, S_{R}^{c})$ the (symmetric) neutrino mass matrix is
\begin{equation}
\mathcal{M}_\nu \;=\;
\begin{pmatrix}
0          & m_D^{\top}      & 0 \\
m_D    & 0          & M^{\top} \\
0          & M          & \mu
\end{pmatrix}.
\label{eq:ISS-mass-matrix}
\end{equation}
For $|\mu|\ll |M|$ and $|m_D|\sim |M|$, the matrix can be diagonalized perturbatively.
Denoting by $O$ the orthogonal matrix that diagonalizes $\mathcal{M}_\nu$,
\begin{equation}
O^{\top}\mathcal{M}_\nu\, O =
\mathrm{diag}\left(m_{\nu}, m_{+}, m_{-}\right),
\end{equation}
with approximate eigenvalues
\begin{align}
 m_{\nu} &\simeq \mu \frac{m_D^{2}}{M^{2}}, \\
 m_{\pm} &\simeq \pm \sqrt{M^{2}+m_D^{2}}
          + \mu\frac{M^{2}}{2\left(M^{2}+m_D^{2}\right)}.
\label{eq:ISS-eigenvalues}
\end{align}

In this one-flavour notation, the coefficient $c_\nu$ denotes the
active-neutrino derivative coupling inherited from the lepton-doublet
operator in Eq.~(\ref{eq:lepton_eft}), while $c_N$ and $c_S$ denote
possible singlet-sector ALP charges. In our notation the ALP interaction with the neutral fermions
$\psi=(\nu,N,S)^T$ can be written as
\begin{align}
    \mathcal{L}_{a\psi\psi}
    &=
    \frac{\partial_\mu a}{f_a}
    \bar{\psi}\gamma^\mu\gamma_5 Q\psi ,
    \qquad
    Q=\mathrm{diag}(c_\nu,c_N,c_S).
\end{align}
After diagonalizing the neutral-fermion mass matrix, this interaction becomes
\begin{align}
    \mathcal{L}_{a\nu\nu}
    &=
    \frac{\partial_\mu a}{f_a}
    \sum_{i,j}
    G_{ij}\,
    \bar{\nu}_i\gamma^\mu\gamma_5\nu_j,
    \qquad
    G=O^TQO ,
    \label{eq:Lapsipsi_derivative}
\end{align}
where $O$ is the matrix that rotates the interaction-basis fields into
mass eigenstates. Equivalently, using the equations of motion, one obtains
\begin{align}
    \mathcal{L}_{a\nu\nu}
    &=
    -i\frac{a}{f_a}
    \sum_{i,j}
    (m_i+m_j)G_{ij}
    \bar{\nu}_i\gamma_5\nu_j .
    \label{eq:Lapsipsi_pseudoscalar}
\end{align}

We translate the constraints on the axion-electron coupling $g_{ae}$ into
constraints on the axion decay constant $f_a$. From
Eq.~(\ref{eq:Lapsipsi_pseudoscalar}), the diagonal coupling to a light
neutrino mass eigenstate is
\begin{equation}
    g_{a\nu_i}
    =
    \frac{2m_i}{f_a}G_{ii}.
\end{equation}
Comparing this expression with the electron coupling in
Eq.~(\ref{eq:gae_def}), we obtain
\begin{equation}
    g_{a\nu_i}
    =
    g_{ae}
    \frac{m_i}{m_e}
    \kappa_i,
    \qquad
    \kappa_i\equiv\frac{G_{ii}}{C_e}.
    \label{eq:gae}
\end{equation}

Here $\kappa_i$ is a model-dependent coefficient determined by the ALP
charge assignments and by the neutral-fermion mixing matrix. In the
phenomenological estimates below we take $\kappa_i=1$ as a benchmark,
rather than as a model-independent prediction. This corresponds to an
effective ALP charge of the light neutrino comparable to the electron
coefficient after rotating to the mass basis. This assumption is not
generic. In particular, in Majoron-like models the relation between
neutrino and charged-lepton couplings depends on the specific seesaw
realization; charged-fermion couplings can be loop-induced and need not be
controlled by the same coefficient as the light-neutrino
coupling~\cite{Heeck_2019,herrerobrocal2024majoroncouplingchargedleptons}.

Therefore, the bounds shown below should be interpreted as benchmark
bounds for $\kappa_i=1$. For other charge assignments they scale as
\begin{equation}
    g_{a\nu_i}\propto \kappa_i,
    \qquad
    \tau\propto g_{a\nu_i}^4\propto \kappa_i^4.
\end{equation}
Since the optical depths computed in Sec.~\ref{sec:flux} are far below
unity already for $\kappa_i=1$, our conclusions remain unchanged for
smaller values of $\kappa_i$.

In what follows we use the shorthand $g_{a\nu}\equiv g_{a\nu_i}$ for a
representative light neutrino eigenstate with $m_i=0.1\,\mathrm{eV}$. Including a normal or inverted neutrino-mass ordering would rescale $g_{a\nu}$ for each eigenstate according to  $g_{a\nu_i}\propto m_i $, but would not change the qualitative conclusions of the estimates below.

\section{Present constraints}
\label{sec:data}
There exist several constraints on the axion-electron coupling $g_{ae}$. 

Direct-detection experiments have provided important laboratory constraints on ALPs. 
XENON100 was among the first detectors to set limits on ALP interactions \cite{Aprile_2014}. 
Subsequent results were obtained (or projected) by experiments such as DARWIN \cite{Aalbers_2016}, EDELWEISS \cite{Armengaud_2018}, and GERDA \cite{Agostini_2020}.

Astrophysical considerations yield some of the most stringent bounds. 
In particular, the evolution of red-giant-branch stars constrains $g_{ae}$ \cite{Capozzi_2020}. 
Additional limits arise from solar physics, including searches based on solar axions and solar-neutrino measurements \cite{Gondolo_2009}.

Additional cosmological constraints could be included; however, in this work we focus only on bounds relevant for ALPs and do not assume a particular role of the ALP in cosmology.

\begin{figure}
% trim=left bottom right top, clip
\centerline{\includegraphics[width=0.8\linewidth]{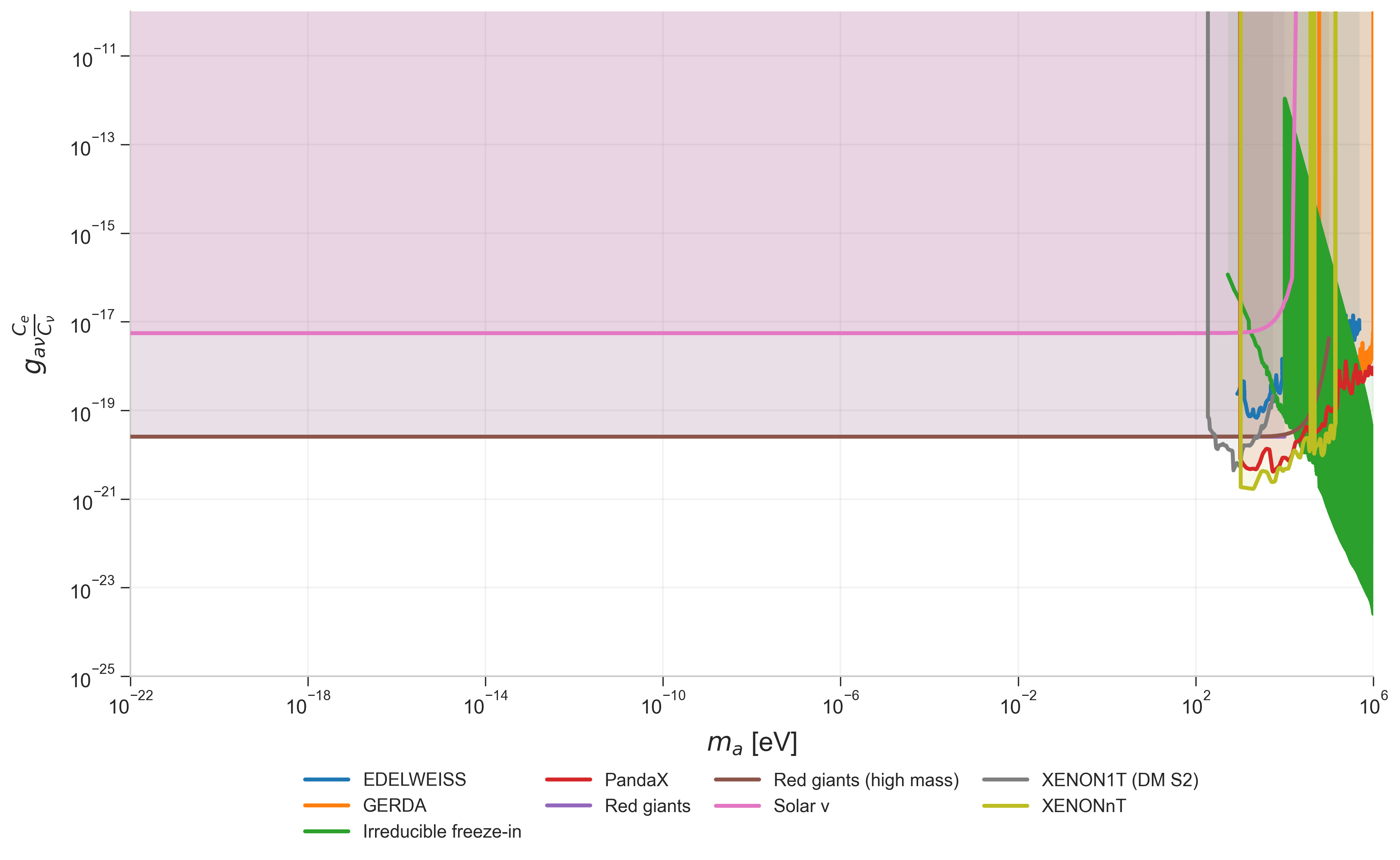}}
\caption{Fig. 1 Benchmark constraints on the axion-neutrino coupling ($g_{a\nu}$) derived from existing axion-electron limits via Eq.\ref{eq:gae}. The translation assumes ($\kappa_i=1$) and ($m_i=0.1 \ \mathrm{eV}$).}
\label{pic1}
\end{figure}

Using Eq.~(\ref{eq:gae}), we obtain the corresponding benchmark
constraints on the neutrino-axion coupling for $\kappa_i=1$, shown in Fig.~\ref{pic1}.

\section{Astrophysical scenarios}
\label{sec:flux}
The axion-neutrino interaction is expected to be extremely weak; nevertheless, it may be probed in environments characterized by high particle densities. We consider two scenarios in which the neutrino-axion interaction could potentially be tested. In the first scenario we explore the  resonant interactions with the C$\nu$B. For the second scenario we test scattering on axion dark matter. 

\subsection{Time delay}
\label{sec:flux:time}
Our first scenario follows Ref.~\cite{Murase:2019xqi} and is based on a time-delay signal induced by neutrino scattering during propagation from a transient source. In particular, we consider scattering of high-energy neutrinos on C$\nu$B neutrinos along the line of sight ($\nu\nu\xrightarrow{}a\xrightarrow{}\nu\nu$). 
In the absence of a direct measurement of the cosmic neutrino background, we adopt the standard $\Lambda$CDM prediction for the relic neutrino distribution. For our order-of-magnitude estimates we treat the C$\nu$B number density as homogeneous and time-independent along the line of sight, using the present-day mean value $n\approx 56  \ \text{cm}^{-3}$ per species. In our estimate we take the cosmic neutrinos to be at rest.
Assuming that the process is resonant, with $s\simeq 2 m_{\nu}\,\epsilon_{\nu}$, where $\epsilon_{\nu}$ is energy of the initial neutrino, the scattering cross section can be written as
\begin{equation}
\begin{split}
\frac{d\sigma_{\nu\nu}}{d\cos\theta}
&=
\frac{g_{a\nu}^4}{32m^2_{\nu}\pi}
\frac{s^2}{(s-m_a^2)^2+m_a^2\Gamma^2} \\
&\quad \times
\left(1+\frac{\epsilon_{\nu}}{m_{\nu}}(1-\cos\theta)\right)^{-2},
\end{split}
\end{equation}
where  $\Gamma=\Gamma_{a\gamma\gamma}+\Gamma_{a\nu\nu}$.
To estimate the impact of this process, we compute the optical depth $\tau = n\,\sigma_{\nu\nu}\,D$, where $D$ is a distance from the source. As a benchmark source we take SN~1987A~\cite{Y_ksel_2007}. We use the largest currently allowed value of $g_{a\nu}$ from Fig.~\ref{pic1} and set the neutrino mass to $m_{\nu}=0.1 \ \text{eV}$. The probability for a neutrino to scatter on a C$\nu$B neutrino along the line of sight is $1-e^{-\tau}$. The optical depth can be estimated as $\tau \approx 10^{-53}$. Hence the overwhelming majority of neutrinos propagate without interacting.

In order to assess whether the effect can become significant, we next consider a second scenario in which the ALP constitutes cold dark matter. 
In this case the relevant target density is \(n_a=\rho_{\rm DM}/m_a\), which is enhanced for smaller ALP masses. 
Therefore, we focus on an ultralight ALP. 
We now turn to a more detailed discussion of this scenario.
\subsection{Axion flux}
\label{sec:flux:axion}
Our second scenario is based on neutrino interactions with axion dark matter. A similar scenario was discussed in~\cite{Reynoso_2022}, illustrating the crucial role of the axion dark-matter density. As a baseline value we take the local Milky Way dark-matter density \\ $\rho_{\text{DM}}=0.4\,\text{GeV}/\text{cm}^3$~\cite{refId0}, assuming that all of the dark matter is composed of axions.

Assuming \(\epsilon_{\nu}\gg m_{\nu}\) and \(m_{a}\ll m_{\nu}\), the neutrino-ALP scattering cross section can be approximated as \cite{Reynoso_2022} 
\[
\sigma_{\nu a} \;\simeq\;
\frac{g_{a\nu}^{4}}{16\pi\,\epsilon_{\nu}\,(m_a+2\epsilon_{\nu})}
\Bigl[2\ln\!\Bigl(\frac{\epsilon_{\nu}}{m_{\nu}}\Bigr)-7\Bigr].
\]

We emphasize that \(n_a\) is highly sensitive to the ALP mass: for ultralight \(m_a\) the number density becomes very large, which maximizes the interaction probability at fixed $(\rho_{\rm DM})$. In this scenario we get $\tau \approx 10^{-15}$. Even though the optical depth in this scenario is significantly larger than in the first scenario, it should be noted that, for ultralight dark matter the axion behaves as a classical field rather than as individual particles~\cite{Berlin_2016}.

\section{Discussion}
\label{sec:dis}
Neither of the two scenarios discussed above yields an observable probe of the axion-neutrino interaction with current sensitivities. We also explored several other scenarios, including scattering on relic neutrinos into three neutrinos via the Fermi interaction, as well as channels involving a neutrino and a photon; in all cases the predicted effects remain unobservable.

We also emphasize that our numerical estimates are obtained for the benchmark choice $\kappa_i=1$.
Since the optical depth scales as
\begin{equation*}
  \tau \propto g_{a\nu_i}^4 \propto \kappa_i^4 ,  
\end{equation*}

different ALP charge assignments can strongly rescale the numerical value of $\tau$.
However, because the benchmark optical depths are already many orders of magnitude below unity, this does not affect our qualitative conclusion for $\kappa_i \sim 1$.

Our estimates  depend on astrophysical and cosmological assumptions. For the time-delay scenario, the C$\nu$B is taken to be homogeneous with a standard $\Lambda$CDM number density; local clustering of relic neutrinos could in principle enhance the effect, but would need to be extremely large to change our qualitative conclusion. For the axion-dark-matter scenario, the result scales with the assumed dark-matter profile and local density; varying $\rho_{\rm DM}$ within commonly used ranges changes $\tau$ only by an overall factor 1 and does not overcome the suppression by many orders of magnitude found above.

A further approximation concerns the ultralight axion limit. When $m_a$ is sufficiently small, the axion dark matter is more appropriately described as a coherent classical field rather than a gas of individual particles. In that regime, the relevant observable may be an oscillation effect induced by a background rather than rare hard scattering events, and the particle-scattering picture used here should be regarded as a conservative order-of-magnitude estimate.

From the perspective of discovery potential, the most promising directions are environments with substantially enhanced target densities like dense dark-matter regions. 

It is possible to decouple the value of $g_{a\gamma\gamma}$ from $f_a$ through the higher-order theories (see e.g. \cite{Rubakov_1997}). It can be done so $f_a$ will take a lower value and make axion-neutrino interaction stronger. We will leave it for future studies.

\section{Conclusions}
\label{sec:concl}
In this paper we studied axion-neutrino interactions in minimal extensions of the Standard Model that generate neutrino masses, focusing on the Type-I seesaw and the inverse seesaw. We then considered astrophysical scenarios in which such interactions could leave an observable effect.

More broadly, in realistic realizations of these neutrino-mass mechanisms the size of
the axion-neutrino coupling is not a freely adjustable parameter: it is tied to the same axion
scale and charge assignments that also control the ALP couplings to charged leptons. As a result,
astrophysical limits on the axion-lepton sector translate into a stringent upper bound on
$g_{a\nu}$, which suppresses any neutrino-propagation signature in the environments considered
here. Therefore, within minimal and phenomenologically viable Type-I and inverse-seesaw models,
none of our benchmark scenarios can yield an observable effect; detectable signals would require
either non-minimal model-building that decouples $g_{a\nu}$ from charged-lepton constraints and
astrophysical settings with drastically enhanced effective target densities.

\section*{Acknowledgments}

The author is very grateful to Sergey Troitsky, Dmitry Kalashnikov, and Dmitry S. Gorbunov for useful discussions.
This work was supported by the Russian Science Foundation, grant no. 22-12-00253-P.

\end{document}